\pdfoutput=1

\documentclass{article}

\usepackage[a4paper, margin=1in]{geometry}
\usepackage[T1]{fontenc}
\usepackage{graphicx}
\usepackage{relsize}
\usepackage{parskip}
\usepackage{microtype}
\usepackage[lighttt]{lmodern}
\usepackage{listings}
\usepackage{tabularx}
\usepackage{caption}
\usepackage{makecell}
\usepackage{hyperref}
\usepackage[
  backend=biber,
  style=numeric,
  sorting=none
]{biblatex}

\hypersetup{
  hidelinks
}

\author{
  \renewcommand{\arraystretch}{1.6}
  \hspace*{-0.3in}
  {\relsize{-1}
  \begin{tabularx}{\linewidth}{XXX}
  Kamalkumar Rathinasamy \newline {\smaller \texttt{kamalkumar\_r@infosys.com}} &
  Jayarama Nettar \newline {\smaller \texttt{jayarama\_nettar@infosys.com}} &
  Amit Kumar \newline {\smaller \texttt{amit\_kumar173@infosys.com}} \\
  Vishal Manchanda \newline {\smaller \texttt{vishal\_manchanda@infosys.com}} &
  Arun Vijayakumar \newline {\smaller \texttt{arun\_vijayakumar@infosys.com}} &
  Ayush Kataria \newline {\smaller \texttt{ayush.kataria@infosys.com}} \\
  Venkateshprasanna Manjunath \newline {\smaller \texttt{venkateshprasanna.m@infosys.com}} &
  Chidambaram GS \newline {\smaller \texttt{chidambaram\_gs@infosys.com}} &
  Jaskirat Singh Sodhi \newline {\smaller \texttt{jaskirat.s@infosys.com}} \\
  Shoeb Shaikh \newline {\smaller \texttt{shoeb.shaikh02@infosys.com}} &
  Wasim Akhtar Khan \newline {\smaller \texttt{wasimakhtar.khan@infosys.com}} &
  Prashant Singh \newline {\smaller \texttt{prashant.singh31@infosys.com}} \\
  Tanishq Dattatray Ige \newline {\smaller \texttt{tanishqdattatray.i@infosys.com}} &
  Vipin Tiwari \newline {\smaller \texttt{vipin.tiwari@infosys.com}} &
  Rajab Ali Mondal \newline {\smaller \texttt{rajab.mondal@infosys.com}} \\
  Harshini K \newline {\smaller \texttt{harshini.k04@infosys.com}} &
  S Reka \newline {\smaller \texttt{s\_reka@infosys.com}} &
  Chetana Amancharla \newline {\smaller \texttt{chetana\_shanbhag@infosys.com}} \\
  Faiz ur Rahman \newline {\smaller \texttt{faiz.rahman@infosys.com}} &
  Harikrishnan P A \newline {\smaller \texttt{harikrishnan.a01@infosys.com}} &
  Indraneel Saha \newline {\smaller \texttt{indraneel\_saha@infosys.com}} \\
  Bhavya Tiwary \newline {\smaller \texttt{bhavya.tiwary@infosys.com}} &
  Navin Shankar Patel \newline {\smaller \texttt{navinshankar\_p@infosys.com}} &
  Pradeep T S \newline {\smaller \texttt{tharmarajan\_p@infosys.com}} \\
  Balaji A J \newline {\smaller \texttt{balaji\_jayaram@infosys.com}} &
  Priyapravas \newline {\smaller \texttt{priyapravas@infosys.com}} &
  Mohammed Rafee Tarafdar \newline {\smaller \texttt{mohammed\_tarafdar@infosys.com}} \\
  \end{tabularx}}
  \\
  \begin{minipage}{\linewidth}
    \vspace{15pt}
    \hspace*{-0.2in}
    \centering
    Infosys Limited
  \end{minipage}
}

\date{}

\title{%
  \rule{\linewidth}{3pt}
  EnterpriseEM:\\Fine-tuned Embeddings for Enterprise Semantic Search
  \rule{\linewidth}{1pt}
}

\newcolumntype{L}{>{\raggedright\arraybackslash\ttfamily\small}X}

\begin{document}

\maketitle

\begin{abstract}
Enterprises grapple with the significant challenge of managing proprietary unstructured data, hindering efficient information retrieval. This has led to the emergence of AI-driven information retrieval solutions, designed to adeptly extract relevant insights to address employee inquiries. These solutions often leverage pre-trained embedding models and generative models as foundational components. While pre-trained embeddings may exhibit proximity or disparity based on their original training objectives, they might not fully align with the unique characteristics of enterprise-specific data, leading to suboptimal alignment with the retrieval goals of enterprise environments. In this paper, we propose a comprehensive methodology for contextualizing pre-trained embedding models to enterprise environments, covering the entire process from data preparation to model fine-tuning and evaluation. By adapting the embeddings to better suit the retrieval tasks prevalent in enterprises, we aim to enhance the performance of information retrieval solutions. We discuss the process of fine-tuning, its effect on retrieval accuracy, and the potential benefits for enterprise information management. Our findings demonstrate the efficacy of fine-tuned embedding models in improving the precision and relevance of search results in enterprise settings.
\end{abstract}

\newpage
\section{Introduction}

In the context of enterprises accumulating proprietary unstructured data, AI-driven information retrieval solutions have emerged as vital tools for extracting relevant answers to employee queries. Traditional methods for developing such solutions often involve choosing between Retrieval Augmented Generation (RAG) or fine-tuned Large Language Models (LLMs). However, fine-tuned LLMs, comprising only generative models, lack a guarantee of factual accuracy, while RAG, comprising an embedding model and a generative model, assures factual precision (Lewis at al., 2020 \cite{Lewis2020}). Despite their superior performance in general, RAG based solutions often rely on pre-trained models, potentially leading to suboptimal alignment with enterprise-specific data.

Addressing this challenge entails exploring two potential avenues: Firstly, recent studies such as RAFT (Zhang et al., 2024 \cite{Zhang2024}) explore the integration of fine-tuned generative models within a RAG pipeline to enhance accuracy, albeit requiring substantial domain-specific data to fine-tune the generative models. Alternatively, leveraging domain-specific embedding models within a RAG pipeline to enhance accuracy remains an underexplored area.

Earlier efforts, such as BioBERT (Lee et al., 2019 \cite{Lee2019}), SciBERT (Beltagy et al., 2019 \cite{Beltagy2019}), and LEGAL-BERT (Chalkidis et al., 2020 \cite{Chalkidis2020}) have effectively demonstrated the efficacy of domain-specific embeddings in information retrieval tasks. These endeavors primarily investigated two methodologies: (a) extending the pre-training of BERT and (b) pre-training BERT from scratch, both employing domain-specific corpora. Despite yielding commendable results, these methodologies necessitated substantial domain-specific corpora, with figures as staggering as 21.3B words for BioBERT, 3.17B tokens for SciBERT, and 11.5GB of text data for LEGAL-BERT, thereby posing significant challenges, particularly in low-resource domains like enterprises. Hence, for such domains, an alternative methodology is to fine-tune an embedding model\footnote{Models such as Sentence-T5 (Ni at al., 2022 \cite{Ni2022}) which produce single vector embeddings might be more suitable over BERT models for retrieval tasks.} with enterprise corpus to arrive at an EnterpriseEM (Enterprise Embedding Model), such as the InfosysEM (Infosys Embedding Model), tailored for enterprise (Infosys Limited) information retrieval tasks. Leveraging an EnterpriseEM in lieu of pre-trained embedding model within a RAG pipeline bolsters the accuracy of semantic search within enterprise contexts.

\section{Datasets}

Our dataset comprised a diverse range of internal Infosys Ltd. data, including technical course contents, internal knowledge base articles, standard operating procedures for technical tasks, a repository of internal technical queries with corresponding resolutions, sales data, and employee blogs. Text data was extracted from various source document formats such as PDF, MS Word, Excel, PowerPoint, and web pages. However, data from audio, video, and image files were not included in this version of the dataset.

\subsection{Extraction}

Langchain's document loaders\footnote{\url{https://python.langchain.com/v0.1/docs/modules/data_connection/document_loaders/}}, including UnstructuredFileLoaders from the community package, provided a convenient abstraction atop renowned parsing libraries, offering comprehensive support for various file formats such as PDFs, MS Word documents, PowerPoint presentations, Excel sheets, HTMLs, markdowns, emails, Evernote files, ODT files, and text files, thereby facilitating efficient data extraction. In addition to this, specialized tools such as PDFMiner\footnote{\url{https://pdfminersix.readthedocs.io/en/latest/}}, PyMuPDF\footnote{\url{https://pymupdf.readthedocs.io/en/latest/}}, python-pptx\footnote{\url{https://python-pptx.readthedocs.io/en/latest/}}, and BeautifulSoup\footnote{\url{https://www.crummy.com/software/BeautifulSoup/}} were utilized for extracting text from PDFs, PowerPoint presentations and HTML files. Non-textual data, such as images and graphics in these files were disregarded during the extraction process.

\subsection{Preprocessing}

The collected data underwent a multi-step preprocessing pipeline to ensure its suitability for synthetic question generation.

\subsubsection*{Masking}

The Presidio Analyzer\footnote{\url{https://microsoft.github.io/presidio/analyzer/}} and Anonymizer\footnote{\url{https://microsoft.github.io/presidio/anonymizer/}} engines were employed to detect and mask personally identifiable information (PII) within text data. Configured with the spaCy model en\_core\_web\_lg\footnote{\url{https://spacy.io/models/en\#en_core_web_lg}} for English language text analysis, the Analyzer Engine identifies potential PII instances, while the Anonymizer Engine masks the identified PII with asterisks. This meticulous approach ensured thorough detection and anonymization of any PII present in the input text, thereby safeguarding individual privacy and ensuring compliance with data protection regulations.

\subsubsection*{Cleaning}

The data cleaning stage addressed various undesirable elements, including XML tags, HTML scripts, and non-ASCII characters. Python libraries, lxml\footnote{\url{https://pypi.org/project/lxml/}} and clean-text\footnote{\url{https://pypi.org/project/clean-text/}}, were employed to effectively remove these noise elements. However, certain noise types, such as tables, lists, and non-English sentences, required manual intervention.

\subsubsection*{Chunking}

The chunking process aimed to segment the cleaned data into contextually relevant units suitable for synthetic question generation. This process involved three key steps:

\begin{itemize}
  \item \textbf{Denormalization of Structured Data:} A subset of our data which was structured, such as tabular or hierarchical data, was denormalized or flattened into plain text, represented as key-value pairs, with the help of a custom python script. The data was then split into chunks using Langchain RecursiveTextSplitter\footnote{\url{https://python.langchain.com/docs/modules/data_connection/document_transformers/recursive_text_splitter/}}, and the important key-value pairs were duplicated in each chunk to preserve the contextual information.
  \item \textbf{Paragraph Separation:} For unstructured data consisting of paragraphs, contextually independent paragraphs were identified and separated from the continuous text stream.
  \item \textbf{Sentence Chunking:} The separated paragraphs were further segmented into smaller chunks using the Langchain RecursiveTextSplitter with appropriate delimiters. The HuggingFace tokenizer BertTokenizerFast was employed, with a maximum chunk size of 512 tokens and an overlap of 40-70 tokens to preserve contextual information across chunks.
\end{itemize}

\begin{figure}[htbp]
    \centering
    \includegraphics[width=\linewidth]{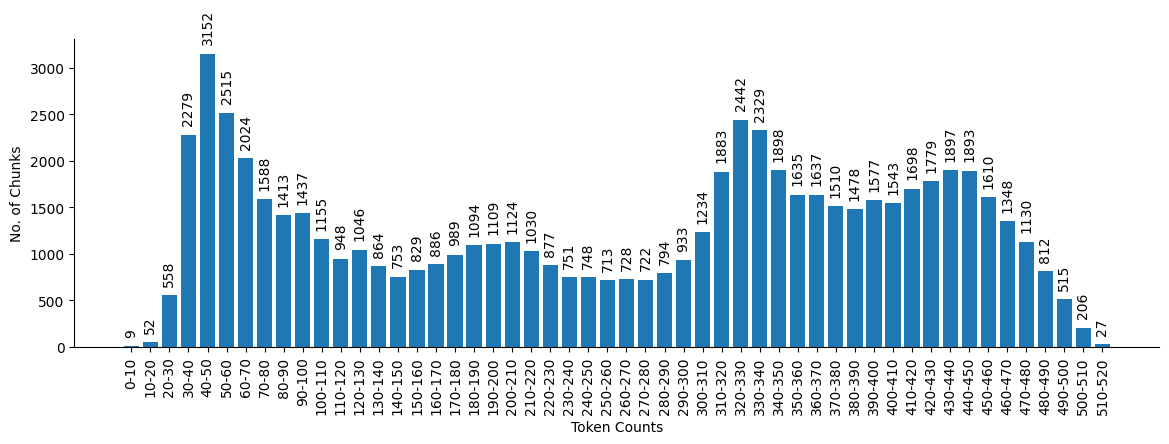}
    \caption{Histogram of the chunk sizes showing the number of chunks with different token counts.}
    \label{fig:chunksize}
\end{figure}

This diverse data collection amounted to approximately 17 million tokens distributed across 65,200 data chunks. The histogram of chunk sizes in our dataset as shown in Figure \ref{fig:chunksize} indicates that most chunks consisted of 300-500 tokens, with many smaller chunks containing less than 100 tokens.

\subsection{Synthetic Question generation}

The final stage involved feeding the cleaned and chunked dataset into a pre-trained large language model (LLM), Mixtral-8x7B-Instruct-v0.1\footnote{\url{https://huggingface.co/mistralai/Mixtral-8x7B-Instruct-v0.1}}, to generate a diverse range of questions (approximately 300,000). These questions encompassed various types, including factual, clarification, interpretation, consequence-related, extractive, subjective, and reasoning-based, achieved using engineered prompts. The synthetic question generation process was carried out in multiple iterations, where the questions generated from each iteration were curated through rigorous review, and substandard questions were discarded to ensure the overall quality of the generated questions.

The following is an example prompt used for generating different types of questions:

\begin{minipage}{\linewidth}
\begin{lstlisting}[columns=fullflexible]
INSTRUCTIONS: Generate a variety of questions spanning different types, including factual, clarification, interpretation, scenario-based, consequence-related, extractive, subjective, reasoning-based, etc., which can be answered using the information from the provided CONTEXT only without any prior knowledge as demonstrated in the EXAMPLE.

CONTEXT: {text}

EXAMPLE:

Example Context: Finacle help banks engage better with their customers, employees, and partners. We do this by helping banks design and deliver truly personalized products and services. Built on a unique engagement hub, our suite helps banks onboard, sell, service, and converse better with customers. In fact, banks running on Finacle have realized an average 19% improvement in their NPS scores.

Example Questions:

1.  Factual: What is the primary focus of Finacle in helping banks?
2.  Clarification: Can you elaborate on how Finacle helps banks engage with their customers, employees, and partners?
3.  Scenario based: Suppose a bank wants to improve its customer engagement. How can Finacle help achieve this goal?
4.  Consequence related: What benefits have banks experienced after implementing Finacle, particularly in terms of their NPS scores?
5.  Extractive: What percentage improvement have banks seen in their NPS scores when using Finacle?
6.  Reasoning based: What might be some reasons behind the effectiveness of Finacle in enhancing customer engagement for banks?
\end{lstlisting}
\end{minipage}

\subsubsection*{Alignment@Scale (Custom Entity Annotation)}

Annotating domain-specific entities in the dataset, synthesizing questions around these annotated entities, and subsequently training the model with these questions facilitates enhanced ranking of answers by prioritizing relevance to specific entities within the inquiry. The domain-specific nature of the annotations contributes to the model's acclimatization with Infosys terminology and subtleties, thereby refining its capacity to grasp contextual nuances within the domain-specific questions, ultimately enhancing the overall performance of the question-answering system.

In pursuit of improving question generation effectiveness and refining the query dataset, initial attempts were made to identify Infosys-specific entities using state-of-the-art libraries like spaCy\footnote{\url{https://spacy.io/api/entityrecognizer}} and NLTK\footnote{\url{https://www.nltk.org/}}, as well as pre-trained entity recognition models. However, these methods exhibited limitations in accurately recognizing Infosys-specific entities. To address this, a novel approach was implemented, combining automated pre-annotation with manual refinement.

A curated dictionary of around 500 organization-wide entities with their types was developed, and a script was created to recognize these entities within the dataset. The entire dataset was then transformed into multiple batches of 50 chunks each. Subsequently, every batch was independently annotated by three annotators trained in manual entity recognition and labelling techniques with a specialized course. The course, created in collaboration with Infosys Education, Training, and Assessment (ETA) department, provided in-depth explanations of entity recognition concepts alongside numerous illustrative examples, equipping annotators with the skills necessary to produce high-quality annotations. Following an automated review process, where only entities mutually agreed by all three annotators in each batch, were incorporated into the final annotated dataset. This process facilitated the creation of a comprehensive high-quality annotated dataset optimized for entity-specific synthetic question generation.

Table \ref{tab:synthetic_question_generation_with_custom_entity_annotation} provides an illustrative example of how engineering the prompt around custom entities can improve the relevancy of generated questions to the domain. In this example, a basic prompt leads to questions that may not fully capture the meaning of the entity ‘Infosys Meridian’, resulting in less relevant questions. However, when the prompt is enhanced with an instruction clarifying that ‘Infosys Meridian’ is a platform, the questions generated demonstrate a clearer understanding of this context. The improved understanding reflected in the generated questions will be passed on to the embedding model during fine-tuning, enhancing its performance on domain-specific tasks.

\begin{table*}[htbp]
  \centering
  \caption{\label{tab:synthetic_question_generation_with_custom_entity_annotation}
    Comparison of synthetic question generation prompts and the corresponding results, with and without custom entity annotations.
  }
  \renewcommand{\arraystretch}{1.5}
  \begin{tabularx}{\textwidth}{p{2cm}XX}
  \Xhline{1pt}
  \textbf{Context:} &
  \multicolumn{2}{>{\hsize=\dimexpr2\hsize+2\tabcolsep+\arrayrulewidth\relax}X}{Infosys and Tennis Australia have joined hands to redefine the tennis experience by leveraging cloud, 3D virtual experiences, AI and mobility. A recent press release further highlights how our enterprise is applying technology to take the game to those virtually who are unable to attend the matches physically. Learn more about the Virtual Hub - Powered by Infosys Meridian - which engages partners of the Australian Open and their guests in premium digital hospitality across the vast expanse of Melbourne Park.} \\
  \Xhline{0.1pt}
  \textbf{Instructions:} &
  Generate 3 questions of different types based on the given context. &
  Generate 3 questions of different types \textbf{around the entity ‘Infosys Meridian’ of type Platform} in the given context. \\
  \Xhline{0.1pt}
  \textbf{Generated Questions:} &
  1. What companies have collaborated to enhance the tennis experience by utilizing cloud, 3D virtual experiences, AI, and mobility?\vspace{5pt}
  
  2. What is the purpose of the Virtual Hub - Powered by Infosys Meridian?\vspace{5pt}
  
  3. How does Infosys' partnership with Tennis Australia demonstrate the application of technology to overcome real-world limitations in sports events? &
  1. What is the name of the platform that Infosys is using to engage partners and their guests digitally at the Australian Open?\vspace{5pt}
  
  2. How does the Infosys Meridian platform contribute to redefining the tennis experience at the Australian Open?\vspace{5pt}
  
  3. Imagine you are a guest of a partner of the Australian Open who cannot attend the event physically. How might the Infosys Meridian platform help you engage with the tournament? \\
  \Xhline{1pt}
  \end{tabularx}
\end{table*}

\subsection{Benchmark Evaluation Dataset}

For benchmark evaluation, we manually curated a dataset comprising 2500 diverse question-chunk pairs from actual end-user queries and corresponding answers provided by internal SMEs, closely mirroring real-world end-user inquiries. This dataset was carefully cleaned and curated to ensure no overlap with the training data and to provide a proportional distribution across various subsets of enterprise data. This dataset was used to assess and compare the accuracy of the models.

\section{Training}

\subsection{Dataset Preparation}

The dataset was split into train and validation datasets at a 95:5 ratio for each type of Infosys data.

\subsection{Foundation Model Selection}

For selecting the foundation models, we considered models that are currently used as foundation models for existing Semantic Search solutions at Infosys. Additionally, we included state-of-the-art models based on HuggingFace MTEB Leaderboard\footnote{\url{https://huggingface.co/spaces/mteb/leaderboard}}, with a preference for open-source models over closed models. Other factors considered during this process included embedding model size, embedding dimensions, context length, and performance on benchmark evaluation datasets. After the selection process, the shortlisted foundation models were evaluated using our benchmark evaluation dataset using BeIR\footnote{\url{https://github.com/beir-cellar/beir}} library. Based on the evaluation results, we selected three models: e5-large-v2\footnote{\url{https://huggingface.co/intfloat/e5-large-v2}, A state-of-the-art general purpose embedding model from E5 family (Wang at al., 2024 \cite{Wang2022}).} for retrieval, mxbai-rerank-large-v1\footnote{\url{https://huggingface.co/mixedbread-ai/mxbai-rerank-large-v1}} for reranking, and colbertv2.0\footnote{\url{https://huggingface.co/colbert-ir/colbertv2.0}, The latest version of ColBERT (Khattab at al., 2020 \cite{Khattab2020}), a contextual late interaction model.} for both retrieval and reranking tasks, which consist of 335M, 435M and 110M parameters respectively.

\subsection{Training Process}

The selected models – e5-large-v2, mxbai-rerank-large-v1, and colbert2.0 – were fine-tuned for retrieval and reranking tasks, optimizing their performance for their respective objectives.

\subsubsection*{Bi-encoder model}

Fine-tuning of the e5-large-v2 foundation model for retrieval tasks was done using the Sentence Transformer\footnote{\url{https://sbert.net/}} library.

Experiments encompassed various training data formats, including [query, positive text], [query, positive text, negative text], and [query, positive text, multiple negative texts] where the negative pairs were identified by hard negative mining technique\footnote{The hard negative mining technique introduced by Wang at al., 2022 \cite{Wang2022a} in their work on Generative Pseudo Labeling for Unsupervised Domain Adaptation of Dense Retrieval, was adopted for hard-mining negative pairs in the training data.} using msmarco-distilbert-base-v3 and msmarco-MiniLM-L-6-v3 models. The [query, positive text] format, which omitted the negative pairs, demonstrated optimal performance.

Multiple experiments with different combinations of hyperparameters were conducted, with the MultipleNegativeRankingLoss function proving most effective. Evaluation utilizing the InformationRetrievalEvaluator class yielded superior results. Optimal hyperparameters included epoch = 3, warmup\_steps = 400, weight\_decay = 0.01, scheduler = WarmupLinear, and learning\_rate = 1e-5.

Fine-tuning was carried out on a single NVIDIA A100 80GB GPU, spanning approximately 8 hours.

\subsubsection*{Cross-encoder model}

For reranking tasks, fine-tuning of the mxbai-rerank-large-v1 model was conducted using the Sentence Transformer library.

Experiments explored various input data formats, including [query, positive text], [query, positive text, negative text], and [query, positive text, multiple negative texts] where the negative pairs were identified by hard negative mining technique using msmarco-distilbert-base-v3\footnote{\url{https://huggingface.co/sentence-transformers/msmarco-distilbert-base-v3}} and msmarco-MiniLM-L-6-v3\footnote{\url{https://huggingface.co/sentence-transformers/msmarco-MiniLM-L-6-v3}} models. The [query, positive text, negative text] format exhibited superior accuracy during evaluation.

Optimal hyperparameters, including weight\_decay = 0.01, scheduler = WarmupLinear, epoch = 3, train\_batch\_size = 16, warmup\_steps = 400, and learning\_rate = 1e-5, were determined utilizing CECorrelationEvaluator.

Fine-tuning was performed on a single NVIDIA A100 80GB GPU, lasting approximately 12 hours.

\subsubsection*{Colbert model}

Fine-tuning of the colbertv2.0 foundation model was accomplished via the RAGatouille\footnote{\url{https://github.com/bclavie/RAGatouille}} library. This model can serve as a standalone retrieval model, leveraging a ColBERT index without requiring further reranking. Alternatively, it can also be used for reranking the results obtained from other retrieval models.

Experiments explored various data formats, with the [query, positive text, negative text] format outperforming others in terms of result accuracy. Negative pairs were generated using the RAGatouille library with the help of bge-small-en-v1.5\footnote{\url{https://huggingface.co/BAAI/bge-small-en-v1.5}} model.

Optimal results were achieved utilizing default RAGatouille hyperparameters, including train\_batch\_size = 32, learning\_rate = 5e-6, and warmup\_steps = 10\% of the total steps (9500 in our case).

Fine-tuning was executed on a single NVIDIA A100 80GB GPU, for about 6 hours.

\section{Evaluation and Results}

A thorough analysis was conducted, exploring different combinations of pre-trained and fine-tuned models across diverse architectures. The evaluation was performed using the BeIR library on our benchmark evaluation dataset, with assessment metrics, including Normalized Discounted Cumulative Gain (NDCG), Mean Average Precision (MAP), Precision, and Recall, computed at a k\_value of @3. These scores together provided a detailed comparison of the model’s performance across different aspects of accuracy. Table \ref{tab:eval_results} summarizes the key findings and performance metrics obtained from this extensive evaluation.

\begin{table}[htbp]
  \centering
  \caption{Summary of Model Configuration and Evaluation Results.}
  \label{tab:eval_results}
  \renewcommand{\arraystretch}{1.5}
  \renewcommand\tabularxcolumn[1]{m{#1}}
    \begin{tabularx}{\textwidth}{c >{\raggedright\arraybackslash}X >{\raggedright\arraybackslash}X c c c c}
    \Xhline{1pt}
      Exp \# & Retrieval Model & Reranking Model & NDCG@3 & MAP@3 & Prec@3 & Rec@3\\
    \Xhline{0.3pt}
      1 & Pre-trained e5-large-v2 & - & 78.7\% & 76.7\% & 84.5\% & 28.2\%\\
    \Xhline{0.1pt}
      2 & Pre-trained e5-large-v2 & Pre-trained mxbai-rerank-large-v1 & 86.6\% & 85.3\% & 90.5\% & 30.2\%\\
    \Xhline{0.1pt}
      3 & Pre-trained e5-large-v2 & Pre-trained colbertv2.0 & 86.5\% & 85.2\% & 90.3\% & 30.1\%\\
    \Xhline{0.1pt}
      4 & Pre-trained colbertv2.0 & - & 90.7\% & 89.2\% & 94.8\% & 31.6\%\\
    \Xhline{0.1pt}
      5 & Pre-trained text-embedding-ada-002 & - & 81.0\% & 78.9\% & 86.8\% & 28.9\%\\
    \Xhline{0.1pt}
      6 & Pre-trained text-embedding-3-small & - & 79.2\% & 77.2\% & 85.2\% & 28.4\%\\
    \Xhline{0.1pt}
      7 & Fine-tuned e5-large-v2 & - & 87.4\% & 85.7\% & 92.2\% & 30.7\%\\
    \Xhline{0.1pt}
      8 & Fine-tuned e5-large-v2 & Pre-trained mxbai-rerank-large-v1 & 90.1\% & 88.7\% & 94.1\% & 31.4\%\\
    \Xhline{0.1pt}
      9 & Fine-tuned e5-large-v2 & Pre-trained colbertv2.0 & 90.4\% & 89.0\% & 94.3\% & 31.4\%\\
    \Xhline{0.1pt}
      10 & Fine-tuned e5-large-v2 & Fine-tuned mxbai-rerank-large-v1 & \textbf{91.2\%} & \textbf{89.9\%} & \textbf{95.1\%} & \textbf{31.7\%}\\
    \Xhline{0.1pt}
      11 & Fine-tuned e5-large-v2 & Fine-tuned colbertv2.0 & \textbf{92.4\%} & \textbf{91.4\%} & \textbf{95.1\%} & \textbf{31.7\%}\\
    \Xhline{0.1pt}
      12 & Fine-tuned colbertv2.0 & - & \textbf{93.4\%} & \textbf{92.4\%} & \textbf{96.2\%} & \textbf{32.1\%}\\
    \Xhline{1pt}
    \end{tabularx}
\end{table}

These findings highlight the versatility of ColBERT models in various stages of the information retrieval pipeline and the efficacy of fine-tuning in enhancing overall information retrieval accuracy. Experiments without re-rankers using only the fine-tuned retrieval models exhibited comparable or even superior performance to their pre-trained counterparts, even when accompanied by pre-trained re-rankers. Overall, these evaluation results reveal that fine-tuned bi-encoder retrieval models, particularly when paired with fine-tuned re-rankers, consistently outperform equivalent pre-trained models by a significant margin, making them best suited for enterprise information retrieval tasks.

Besides the improvements in accuracy, an EnterpriseEM offers the advantage of consolidating multiple embedding models into a single model, simplifying the hosting and serving processes.

\section{Conclusion and Future Work}

In conclusion, this paper has proposed a methodology aimed at enhancing the performance of information retrieval solutions within enterprise environments by fine-tuning pre-trained embedding models. By addressing the challenge of efficiently managing proprietary unstructured data, our approach offers promising prospects for improving the precision and relevance of search results. Particularly when coupled with our techniques for synthetic question generation based on annotated custom entities, our methodology demonstrates its potential to adapt embeddings to better suit prevalent information retrieval tasks in enterprises, thereby facilitating more efficient knowledge discovery and decision-making processes.

Moving forward, future research in the field of fine-tuned embedding models for information retrieval tasks could explore emerging techniques and strategies such as semantic chunking\footnote{\url{https://python.langchain.com/docs/modules/data_connection/document_transformers/semantic-chunker/}}, which aims to improve the semantic search accuracy by considering the semantic similarity of sentences within chunks. Moreover, while this paper focuses on enterprise environments, there is potential to generalize and apply our approach to other domains. Future work will consider adapting and testing the methodology for various domains, which may involve making adjustments to fit the specific characteristics and requirements of those areas.

Additionally, exploring techniques for multimedia data extraction presents an opportunity to maximize the utilization of available enterprise data, enriching the training corpus and potentially improving model performance in capturing diverse types of information. Continuous integration and adaptation of state-of-the-art models and methodologies will always remain crucial, ensuring embedding models evolve to meet the evolving needs of enterprise information retrieval tasks. By pursuing these avenues for future exploration, we aim to further enhance the effectiveness and applicability of fine-tuned embedding models in enterprise information retrieval tasks.

\printbibliography

\end{document}